\documentclass[fleqn,usenatbib]{mnras}

\usepackage{newtxtext,newtxmath}
\usepackage[T1]{fontenc}

\DeclareRobustCommand{\VAN}[3]{#2}
\let\VANthebibliography\thebibliography
\def\thebibliography{\DeclareRobustCommand{\VAN}[3]{##3}\VANthebibliography}

\usepackage{graphicx}	% Including figure files
\usepackage{amsmath}	% Advanced maths commands
\usepackage{orcidlink}
\usepackage{xcolor} % required for colored text
\usepackage{ulem} % required for striking out text with sout
\usepackage{booktabs}

\newcommand{\revision}[1]{#1}
\title[Formation of rotating supergiants via stellar mergers in dense clusters]{\vspace{-0.75cm} Formation of rotating supergiants via stellar mergers in dense clusters: \\ Implications for black hole natal spins}

\author[Satish et al.]{
\parbox[t]{\textwidth}{\vspace{-0.75cm}
Ishaan Satish\orcidlink{0009-0002-0758-6983},
$^{1, \star}$
Kyle Kremer\orcidlink{0000-0002-4086-3180},
$^{1}$
Fulya K{\i}ro\u{g}lu\orcidlink{0000-0003-4412-2176},
$^{2}$
Daichi Tsuna\orcidlink{0000-0002-6347-3089},
$^{3}$
and Shahed Shayan Arani\orcidlink{0009-0005-4168-2858}
$^{4,5}$} \vspace{-0.25cm} \\
$^{1}$University of California, San Diego, Department of Astronomy \& Astrophysics; La Jolla, CA 92093, USA\\
$^{2}$Center for Interdisciplinary Exploration \& Research in Astrophysics (CIERA), Northwestern University, Evanston, IL 60201, USA\\
$^{3}$Center for Astrophysics $|$ Harvard \& Smithsonian, 60 Garden St, Cambridge, MA 02138, USA\\
$^{4}$University of California, San Diego, Department of Physics; La Jolla, CA 92093, USA\\
$^{5}$Center for Dark Cosmology and Gravitation, University of California, San Diego, La Jolla, CA 92093-0319\\
$^\star$E-mail: ishaansatish04@gmail.com
\vspace{-0.5cm}
}

\date{\vspace{-0.5cm}\today}
\pubyear{2026}
\begin{document}
\maketitle

\begin{abstract}
\revision{We investigate how massive stellar mergers in young star clusters imprint on black hole spin distributions and the implications for gravitational wave sources. While mechanisms like hierarchical mergers and binary accretion can form rapidly spinning black holes, our central hypothesis is that angular momentum transferred during stellar mergers substantially affects the spins of the merger products and resulting black holes, with some merger products evolving into collapsar-like objects that retain thick accretion disks enabling efficient spin up. This is in contrast to the general expectation that black holes form with small spins, having shed most of their envelope angular momentum via winds and expansion before core collapse.
Using roughly 150 N-body models generated with the \texttt{Cluster Monte Carlo} code, \texttt{CMC}, we analyze stellar mergers that lead to black hole formation, prioritizing ``significant'' events with mass ratio $q>0.1$. After identifying optimal candidates, we explore stellar structure and post-merger evolution implications with MESA models to capture angular momentum injection and pre-collapse profiles most relevant for BH natal spin. In our current dataset representative of Milky Way-like globular clusters, up to roughly half of black holes are formed from such mergers, with roughly $10\%$ from ``significant'' mergers. Angular momentum estimates indicate substantial spin-up during the merger, and trends with mass ratio and stellar properties suggest strong correlations with the final black hole spin. In some cases, dimensionless spin parameters of $a\simeq 0.5$ or more are expected. This process has important implications for the dynamical formation and retention of gravitational wave sources in clusters.}
\end{abstract}
\section{Introduction} 
\label{sec:introduction}

Dense stellar clusters offer rich environments for studying an array of dynamical processes, in particular close gravitational encounters that lead to stellar interactions \citep[e.g.,][]{HeggieHut2003}. Frequent close passages can result in tidal captures \citep[e.g.,][]{Fabian1975}, tidal disruptions \citep[e.g.,][]{Rees1988}, and even direct physical collisions \citep[e.g.,][]{BenzHills1987,Fregeau2004}. In old globular clusters, stellar collisions are understood to be a key formation mechanism for blue straggler stars \citep[e.g,][]{Sandage1953,Bailyn1995,Ferraro2012}. In young clusters that still host significant populations of massive stars, stellar collisions can form objects with exotic properties, such as very massive, rapidly rotating, and/or highly magnetized stars \citep[e.g.,][]{PortegiesZwart1999,PortegiesZwartMcMillan2002,Glebbeek2009}, which can ultimately impact the properties of the black holes (BHs) formed.

For isolated stars and binaries, physical processes such as pre-collapse line-driven winds \citep[e.g.,][]{Vink2001}, angular momentum transport \citep[e.g.,][]{Heger2000}, explosion mechanism \citep[e.g.,][]{Fryer2012}, pair instabilities \citep[e.g.,][]{WoosleyHeger2021}, and post-collapse fallback/accretion \citep[e.g.,][]{WoosleyWeaver1995,MacFadyen2001,Zhang2008} all leave distinct fingerprints upon the masses and spins of the BHs formed. State-of-the-art numerical simulations of core-collapse supernovae reveal a complex and diverse landscape \citep[for a recent review, see][and references therein]{BurrowsVartanyan2021}. In young clusters, massive stellar collisions may add further complexity, potentially producing BH remnants with properties inaccessible to isolated stellar evolution channels \citep[e.g.,][]{DiCarlo2019,Kremer2020,Gonzalez2021,Kiroglu2025}. In extreme cases, collisions may occur in a runaway scenario that may provide a pathway for seeding massive BHs \citep[e.g.,][]{PortegiesZwart2004,Gurkan2006,GonzalezPrieto2024}. Stellar clusters therefore provide natural laboratories to study how stellar collisions shape the BH mass and spin distributions relative to isolated stellar evolution.

In the past decade, the LIGO--Virgo--KAGRA (LVK) network of gravitational wave detectors have provided tremendous new insight into stellar-mass BH populations \citep[e.g.,][]{LIGO2016,LIGO2025}. Although the precise astrophysical pathways through which LVK sources form remain uncertain \citep[e.g.,][]{Zevin2021,MandelBroekgaarden2022}, a general consensus has emerged that dynamical formation within dense stellar clusters plays a key role \citep[e.g.,][]{PortegiesZwartMcMillan2000,Rodriguez2016mergers,AntoniniRasio2016,Askar2017,Banerjee2017,Samsing2018,Fragione2019,kremer2020modeling,Mapelli2022,Ye2025,Agrawal2026}. The current population of LVK sources offer several key clues that hint at dynamical origin. For instance, the presence of massive BHs in excess of $50\,M_{\odot}$ that occupy the ``upper mass gap'' expected to be subject to the pair instability; LVK events GW190521 \citep{GW190521} and GW231123 \citep{GW231123} are flagship examples. In dense star clusters, massive BHs within or beyond the pair instability mass gap can naturally form via hierarchical mergers where the remnants of previous mergers pair up with new BHs and merge again \citep[e.g.,][]{MillerHamilton2002,Fishbach2017,GerosaBerti2017,Rodriguez2019,Antonini2019,FragioneRasio2023}. BHs born via previous mergers (often referred to as ``second generation'' BHs) are predicted to have high spins (dimensionless spin parameter $a\approx 0.7$ for initially non-spinning components) as a result of the merger dynamics \citep[e.g.,][]{Rezzolla2008,BertiVolonteri2008}. The recent LVK events GW241011 and GW241110, which both feature rapid and well-measured primary spins, non-negligible spin-orbit misalignment, and unequal mass ratios, provide perhaps the most direct evidence yet for hierarchical origin \citep{GW241011}.

The formation of hierarchical mergers in dense star cluster relies critically on retention of the initial merger product within the host cluster. Retention is inhibited by large gravitational wave recoil kicks attained by BHs upon merger, arising from asymmetries in the gravitational wave emission from the binary mass ratio and component spins \citep[e.g.,][]{Bekenstein1973,Favata2004,Campanelli2007,Lousto2008}. In lower-mass clusters, the occurrence of hierarchical mergers requires the majority of first-generation BHs (the BHs formed via stellar collapse) to be born with small natal spins, $a\approx 0$; higher spins of $a\approx 0.2$ and higher suppress significantly the retention of first-generation BH merger products \citep{Rodriguez2019}. In this case, a detailed understanding of the natal spin values of BHs in these environments, taking into account stellar interactions enabled by dynamics, is crucial for predicting the corresponding rate of hierarchical mergers \citep[e.g.,][]{Kiroglu2025}.

Standard stellar evolution theory strongly favors the production of slowly rotating BHs \citep[e.g.,][]{FullerMa2019}. In typical single-star evolution, as a massive star expands into a supergiant, internal angular momentum is efficiently transported from the core to the extended envelope, inhibiting spin of the BH attained when the core ultimately collapses. Of course, there are ways to circumvent this process. In a subset of massive stars, the envelope may collapse into a disk that subsequently accretes onto the central BH and spins it up \citep[e.g.,][]{MacFadyen2001}. Binary interactions can also significantly alter the spin. BHs in high-mass X-ray binaries are observed to have high spins approaching $a\approx 1$, presumably attained via accretion from a binary companion \citep[e.g.,][]{Gou2011, Qin2019}. For close binaries, chemically-homogeneous evolution incited by tidal interactions can lead to retention of angular momentum and formation of highly-spinning BH binaries \citep[e.g.,][]{MandelDeMink2016, Marchant2016}. \revision{Furthermore, a black hole progenitor can be tidally spun up by a companion star prior to collapse, leading to the formation of a rapidly spinning black hole \citep[e.g.,][]{Kushnir2016, Hotokezaka2017, Wu2025}.} Pre-collapse stellar mergers can also produce overmassive stellar envelopes that yield relatively compact blue supergiants \citep[BSGs; e.g.,][]{Spera2019, DiCarlo2019}. These objects remain relatively compact up to collapse, enabling retention of angular momentum compared to more extended red supergiants (RSGs), ultimately leading to more highly-spinning BHs \citep{Tsuna2025}.

The high densities in stellar clusters may enhance further the rate of spinning BHs via all types of stellar interactions \citep[e.g.,][]{Kiroglu2025a}. Further, dense clusters provide pathways for any BHs born with non-zero natal spins to later pair up into binaries and merge. This in turn carries important implications for gravitational wave recoil kicks, BH retention, and the formation of hierarchical mergers in these dense environments. In this paper, we explore how stellar collisions and mergers that occur prior to stellar collapse may imprint themselves on the spins of BHs formed in clusters, and ultimately influence the retention of binary BH merger products. We use cluster N-body models computed using the \texttt{Cluster Monte Carlo} code \texttt{CMC} to explore the basic mechanism outlined in \citet{Tsuna2025} where a stellar merger leads to formation of blue supergiant and ultimately a moderate to highly-spinning BH. We then track the subsequent evolution of BHs formed via stellar collisions in our models, and explore how higher natal spins may impact long term retention from gravitational wave recoil kicks.

This paper is organized as follows: In Section~\ref{sec:methods}, we describe our numerical methods, including the \texttt{CMC} modeling framework, our data extraction pipeline, and the grid of cluster models used to investigate varying initial conditions. In Section~\ref{sec:CMC_catalog}, we present our results from the \texttt{CMC Cluster Catalog} \citep{kremer2020modeling}, characterizing the demographics of stellar collisions and identifying the subset of interaction products that ultimately form BHs. Section~\ref{sec:theoretical_uncertainties} explores theoretical uncertainties in initial cluster properties---specifically primordial binary fractions, mass segregation, and natal kick prescriptions---and how these factors enhance the formation and retention of spinning BHs. In Section~\ref{sec:GWrecoil}, we discuss the implications for gravitational-wave recoil kicks, utilizing our updated spin predictions to evaluate the likelihood of merger product retention in dense clusters. Finally, we summarize our key findings and provide concluding remarks in Section~\ref{sec:conclusions}.

\section{Modeling stellar clusters with \texttt{CMC}} 
\label{sec:methods}

To study cluster evolution, we use the \texttt{Cluster Monte Carlo} code \texttt{CMC} \citep{Rodriguez2022}. \texttt{CMC} is a H\'{e}non-type Monte Carlo code for stellar dynamics that includes a number of physical processes including two-body relaxation, stellar collisions and mergers, stellar and binary evolution using the population synthesis code \texttt{COSMIC} \citep{Breivik2020}, direct integration of small $N$ resonant encounters using the \texttt{fewbody} package \citep{Fregeau2004}, post-Newtonian corrections to all BH dynamical encounters \citep{Rodriguez2018}, and gravitational wave recoil kicks for all BH mergers. \texttt{CMC} effectively reproduces dynamical evolution of typical globular clusters over their full lifetimes, matching well the present day properties of globular clusters in the local Universe \citep{kremer2020modeling,Rui2021}, as well as a variety of compact object sources including millisecond pulsars \citep[e.g.,][]{Ye2019}, X-ray binaries \citep[e.g.,][]{Kremer2018}, detached BH binaries \citep[e.g.,][]{Kremer2018_ngc3201}, white dwarfs \citep[e.g.,][]{Kremer2021b}, and binary BH (BBH) mergers \citep[e.g.,][]{Rodriguez2016}.

\texttt{CMC} tracks three types of close stellar interactions: (i) ``collisions" that result from direct physical contact during either single-single or binary-mediated fewbody encounters \citep{Fregeau2007}; (ii) ``mergers'' that result from unstable mass transfer in stellar binaries (i.e., failed common envelopes). Mergers may result from primordial binaries in the cluster, or binaries that are dynamically assembled and later enter Roche contact; and (iii) ``stable mass transfer'' during Roche lobe overflow in a binary. In this study, we focus on the former two types of interactions, and do not consider stable mass transfer. In globular clusters between $10^5-10^6 M_{\odot}$, typical velocity dispersions ($\lesssim 10\, $km/s) are significantly less than the escape velocity of individual stars, making the distinction between dynamically-mediated collisions and binary-mediated mergers largely semantic \citep[this is not true for more massive nuclear clusters where the relative velocity of colliding objects can exceed $100\,$km/s or more; e.g.,][]{Rose2026}. For the purposes of this study, we treat the outcome of collisions and mergers identically but retain the distinction between these two terms to clarify the mechanism through which the interaction occurs. We also use ``stellar interaction'' as a broad term throughout to refer collectively to both collisions and mergers, when the distinction between the two does not matter.

We employ two separate sets of \texttt{CMC} models, each described below.

\subsection{\texttt{CMC Cluster Catalog}}

As our default set of models, we utilize the \texttt{CMC Cluster Catalog} \citep{kremer2020modeling}. This suite of 148 models spans a range of cluster properties including initial number of objects ($N=[2,4,8,16,32]\times10^5$), initial virial radius ($r_v=[0.5,1,2,4]\,$pc), metallicity ($Z=[0.01,0.1,1.0]Z_\odot$), and galactocentric distance ($R_{\rm gc}=[2,8,20]\,$kpc) within a Milky Way-like potential. Collectively this set of models spans the full parameter space for Milky Way globular clusters at present.

We refer the reader to \citet{kremer2020modeling} for a detailed description of all assumptions for these models, and summarize here the key points relevant to massive stars and BH dynamics. All models are initialized with a \citet{Kroupa2001} stellar mass function ranging from $0.08-150\,M_{\odot}$. We assume an initial binary fraction of $5\%$ across all stellar masses, reflective of low binary fractions observed in present-day globular clusters \citep[e.g.,][]{Milone2012}. We assume all stellar collisions occur in the ``sticky sphere'' limit where zero mass is lost \citep[e.g.,][]{Rose2026,GonzalezPrieto2026}. We use the metallicity-dependent stellar wind prescriptions from \citet{Vink2001}. These play a crucial role in the mass function of BHs across our full range of metallicities. We follow the ``rapid'' model of \citet{Fryer2012} to compute BH masses formed via core collapse supernovae, and adopt the pair instability prescriptions from \citet{Belczynski2016}. BH natal kicks are computed by sampling from the \citet{Hobbs2005} distribution, and scaling down according to the fractional mass of fallback material computed as in \citet{Fryer2012}. Following \citet{FullerMa2019}, we assume all BHs formed via stellar collapse are born with zero natal spin. We explore in post-processing how this natal spin assumption may change for BHs that are formed from stellar collision or merger products. 

\subsection{New models to explore effect of initial cluster properties on stellar interactions}

Although the \texttt{CMC Catalog Models} reproduce well the \textit{present-day} properties of Galactic clusters, which are quite well characterized, the initial properties of globular cluster progenitors remain highly uncertain; properties at late times do not necessarily reflect birth conditions, particularly for massive stars. A number of these initial conditions can impact the rate of massive stellar interactions at early times, and therefore, the prospects for altering the masses and spins of the BHs formed. To explore these uncertainties, we supplement the \texttt{CMC Cluster Catalog} models with a new set of models that vary the following parameters:

\textit{Initial binary fraction for massive stars:} Observational surveys of young massive clusters and star-forming regions indicate that the vast majority of massive stars are born in stellar multiples \citep[e.g.,][]{Offner2023,Sana2025}. Higher binary fractions lead to significantly higher collision rates by expanding the effective interaction cross-section from the physical radii of individual stars to the significantly larger orbital scales of binary systems \citep[e.g.,][]{Fregeau2004}. To this point, recent \texttt{CMC} simulations with binary fractions up to $100\%$ for massive stars feature markedly enhanced interaction rates, and may play an important role in the onset of collisional runaways and formation of BHs in the upper-mass gap and above \citep[e.g.,][]{Gonzalez2021,Kiroglu2025a,Kiroglu2025}. \revision{Additionally, initial orbital eccentricities for all binary systems are drawn from a thermal eccentricity distribution}.

To explore this effect, we increase the binary fraction to $100\%$ for stars with initial masses of $15\,M_{\odot}$ or more. For these massive binaries, we draw initial mass ratios in the range $q=[0.6,1]$ and orbital periods from the distribution $dn/d \log P \propto P^{-0.55}$ \citep[e.g.,][]{Chatterjee2017_theoretical} from contact to the local hard-soft boundary.

\textit{Primordial mass segregation:} Observations of young massive clusters (ages of roughly $10\,$Myr or less) reveal an increased concentration of massive stars toward their centers \citep[e.g.,][]{Hillenbrand1998,Fischer1998,Gouliermis2004,Stolte2006}, which can affect the rate of stellar interactions at early times \citep[e.g.,][]{Gurkan2006,Ardi2008,Goswami2012,Kremer2020}. To explore this, we run a subset of models with primordial mass segregation, following the prescription of \citet{Kremer2020}, which in turn follows \citet{Baumgardt2008}. For a fixed number density profile, stars are sorted such that the most massive stars have, on average, the lowest specific energy, and are thus, on average, closer to the cluster center.

\textit{BH fallback prescription and natal kicks:} Stellar-mass BHs play a significant role in the long-term dynamics of their host cluster \citep[e.g.,][]{Kremer2026}. The number of BHs retained following formation via stellar collapse, as well as their mass function, depends on the fallback and natal kick prescriptions assumed. This in turn impacts the number of BBH mergers produced \citep[e.g.,][]{Chatterjee2017,Agrawal2026}. As an alternative to the \texttt{CMC Cluster Catalog} defaults, we also run a set of models adopting the ``delayed'' fallback prescription from \citet{Fryer2012} \revision{rather than the ``rapid'' mechanism. This choice is motivated both by recent gravitational-wave observations from LIGO indicating that the delayed mechanism may be more physically realistic, as well as our interest in thoroughly exploring the parameter space. In these models we} also assume (as an extreme upper limit) that BHs form with zero natal kicks, such that all BHs are retained upon formation. In practice, this increases the number of BHs retained by roughly $50\%$ compared to our default models using the ``rapid'' fallback and kick model. By artificially forcing retention of these BHs, we can study the subsequent changes in the BBH formation rate and the downstream effects on the gravitational-wave recoil kicks these systems receive upon merging.

\begin{figure*}
    \centering
    \includegraphics[width=0.95\linewidth]{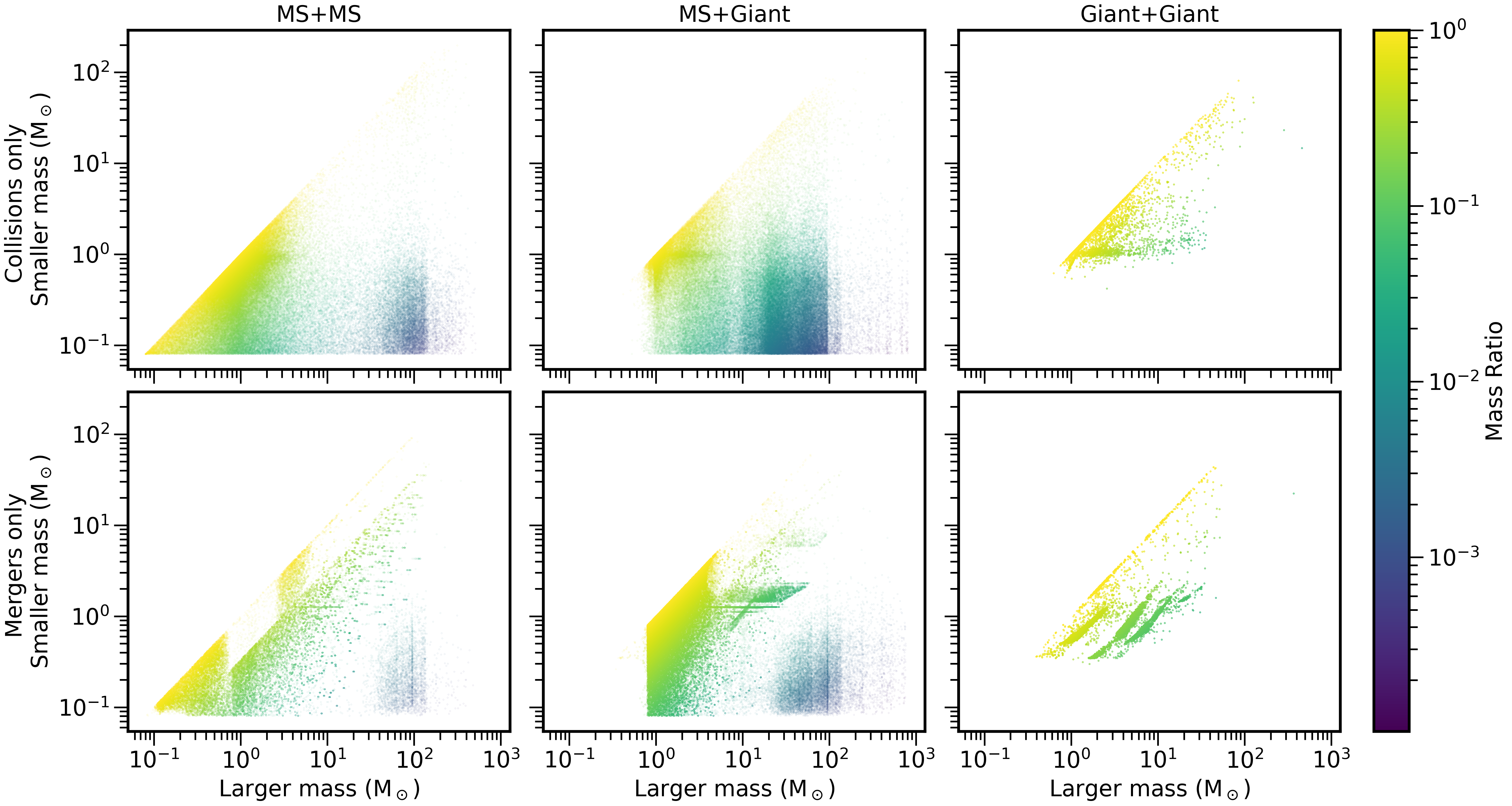}
    \caption{Distribution of stellar interaction mass ratios from the \texttt{CMC Cluster Catalog} simulations, with the larger stellar mass on the x-axis and the smaller mass on the y-axis. Plots are separated by interaction type, from left to right: Main sequence + main sequence (MS+MS), MS+giant, and giant–giant. Upper (lower) panels show stellar collisions (mergers).}
    \label{fig:all_collisions}
\end{figure*}

\section{Results for A Milky Way Cluster Sample} 
\label{sec:CMC_catalog}

As a first step, we begin with the results from the \texttt{CMC Cluster Catalog}. In Section~\ref{sec:CMCcatalog_all_collisions} we describe the full set of collisions and mergers (regardless of mass), and then (Section~\ref{sec:BH_formation}) explore the high-mass collision events that lead to BH formation. In Section~\ref{sec:CMCcatalog_spins}, we describe implications for BH spins. The results of this section can be interpreted as a reasonable proxy for a Milky Way-like globular cluster population. 

\subsection{Stellar collisions and mergers of all masses}
\label{sec:CMCcatalog_all_collisions}
%%%%%%%%%%%%%%%%%%%%%%%%%%%%%%%%%%%%%%%%%%%%%%%%%%%%%%%%%%%%%%%%%%%%%%%%%%%%%%%%%%%%%%%%%%%

Figure~\ref{fig:all_collisions} shows all stellar interactions occurring in the \texttt{CMC Cluster Catalog} models. Here we separate stellar collisions (top panels) and stellar mergers (bottom panels), using the distinctions described in Section~\ref{sec:methods}. We further classify stellar interactions by the types of stars involved. From left to right, we show results for main sequence + main sequence stars (MS+MS), MS+giant, and giant + giant. \revision{Note that in our categorization, the term `giant' broadly encompasses all post-main-sequence evolved stars, including stripped helium stars. The inclusion of these stripped stars accounts for the highly-unequal-mass mergers visible in the giant+giant panels. None of our final candidates of interest involved one of these stripped helium stars.} 
In total (across all models), we identify $148,851$ ($79,752$) MS+MS collisions (mergers), $122,988$ ($438,530$) MS+giant collisions (mergers), and $3,260$ ($8,831$) giant+giant collisions (mergers). 

First consider stellar collisions (top panels): For MS+MS and giant+giant collisions, the most common outcome is a collision between two stars of comparable mass; $48.95\%$ ($76.87\%$) of all collisions for MS+MS (giant+giant) have mass ratio $q > 0.5$, where $q=M_2/M_1$ and $M_2$ and $M_1$ are the secondary and primary masses. This bias toward equal mass events is expected for collisions of two giants, simply because all stars are initialized at the same in our simulations and therefore similar masses evolve off the main sequence at similar times. The trend toward equal mass events for MS+MS collisions arises from the collision cross section and the stellar mass function. In the gravitational-focusing limit, the rate between collisions of stellar species $[M_1,R_1]$ and $[M_2,R_2]$, where $M_i$ is mass and $R_i$ is radius, is given by:

\begin{equation}
\label{eq:gamma}
    \Gamma_{\mathrm{coll}} = n_2 \pi r_p^2 v_\infty \left( 1 + \frac{2G(M_1 + M_2)}{r_p v_\infty^2} \right) N_1,
\end{equation}
where $n_2$ denotes the number density of the target stars, $r_p = R_1 + R_2$ signifies the closest pericenter distance required for a collision to occur, $v_\infty$ represents the relative velocity of the stellar pair at infinity (typically assumed to be the velocity dispersion of the host cluster), and $N_1$ is the total population of stars with the primary mass $M_1$. The \citet{Kroupa2001} mass function scales as $dN/dm\propto m^{-2.3}$ for $m>\revision{0.5}M_\odot$. Stellar radii scale with mass as roughly $R\propto m^{0.6}$ for MS stars. Thus, although more massive MS stars have larger radii and cross sections for collision, the fact that they are much more rare (and also have far shorter MS lifetimes) dominates the collision rate. As a consequence, low-mass MS stars are most likely to merge with other low-mass stars. This basic trend also explains why MS+giant collisions (middle panel) most commonly involve low-mass MS stars; only $14.8\%$ of all MS+giant collisions have mass ratio $q >0.5$. For these events, the large radius of the giant maximizes the collision cross section while the large number of low-mass MS stars maximizes the target population. 

Next, consider stellar mergers (bottom panels).
Similar to MS+MS collisions, MS+MS mergers mostly commonly involve near equal mass pairings; 84.7\% of all MS+MS mergers have mass ratio $q > 0.1$. This is mainly because our primordial binary scheme preferentially pairs stars of comparable mass (Section~\ref{sec:methods}). The merger distributions contains a clear feature at $q\approx 0.3$, arising from our assumption that dynamically unstable mass transfer occurs at this critical mass ratio (\citet{Hurley2002}, but note this is highly uncertain \citep[e.g.,][]{Belczynski2008, Claeys2014}.
Unlike MS+giant collisions, MS+giant mergers most commonly involve near-equal mass components; 84.2\% of all MS+giant mergers have $q > 0.1$. Where the collision rate is dominated by the increased cross section for more massive giants (and the steep IMF), the merger rate is dominated by the fact that comparable mass stars are preferentially paired at birth.

\begin{figure*}
    \centering
    \includegraphics[width=1.0\linewidth]{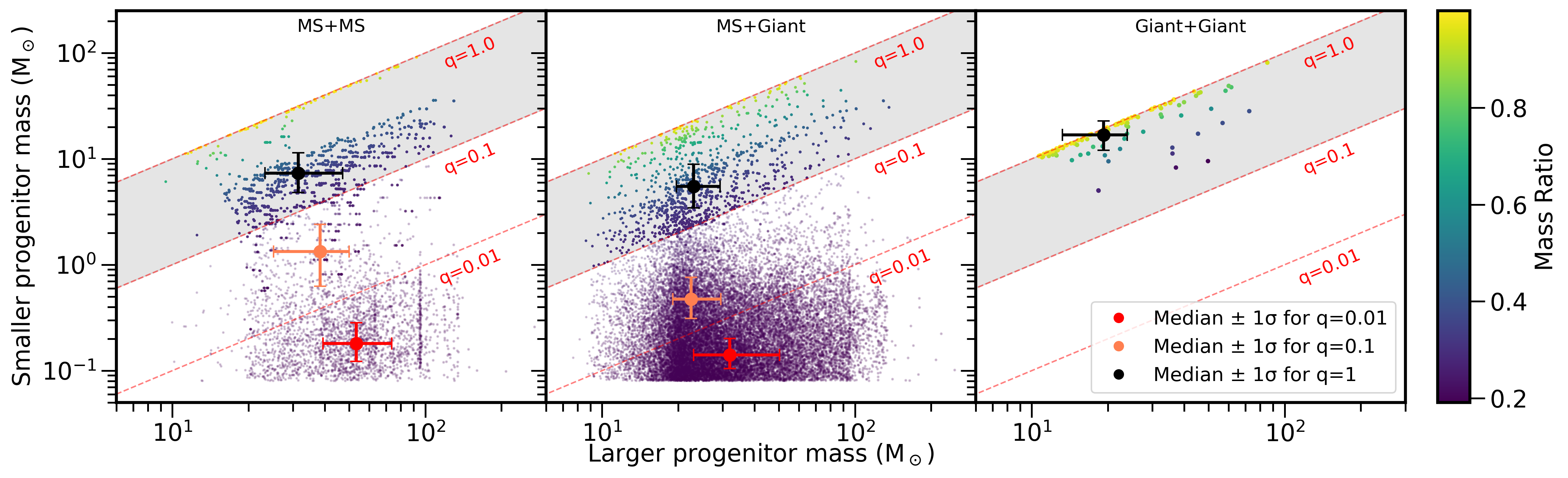}
    \caption{Mass ratio distributions of stellar interactions that form BHs, separated by stellar type. Dotted lines mark $q= [0.01,0.1,1]$; interactions with $q\geq 0.1$ are classified as ``dynamically significant'' and are shaded in gray.}
    \label{fig:BH_collisions} 
\end{figure*}

\subsection{Black holes formed via stellar interaction products}
\label{sec:BH_formation}

Next, we restrict the analysis to only those collisions/mergers that ultimately collapse to form BHs. In practice, this restricts us to only the most massive events in Figure~\ref{fig:all_collisions}. Figure~\ref{fig:BH_collisions} shows the mass pairs for all such BH-forming interactions, again separated by interaction type. Since $74.5\%$ of events occur via collisions (as opposed to mergers), we do not distinguish here between the merger and collision channels.
For each region, the median collision and associated 1$\sigma$ are marked by the red, orange, and black points, respectively. Collisions with $q \geq 0.1$ are designated as ``significant,” in the sense that they are most likely to impart non-negligible angular momentum to the merger product.

\begin{figure*}
    \centering
    \includegraphics[width=0.95\linewidth]{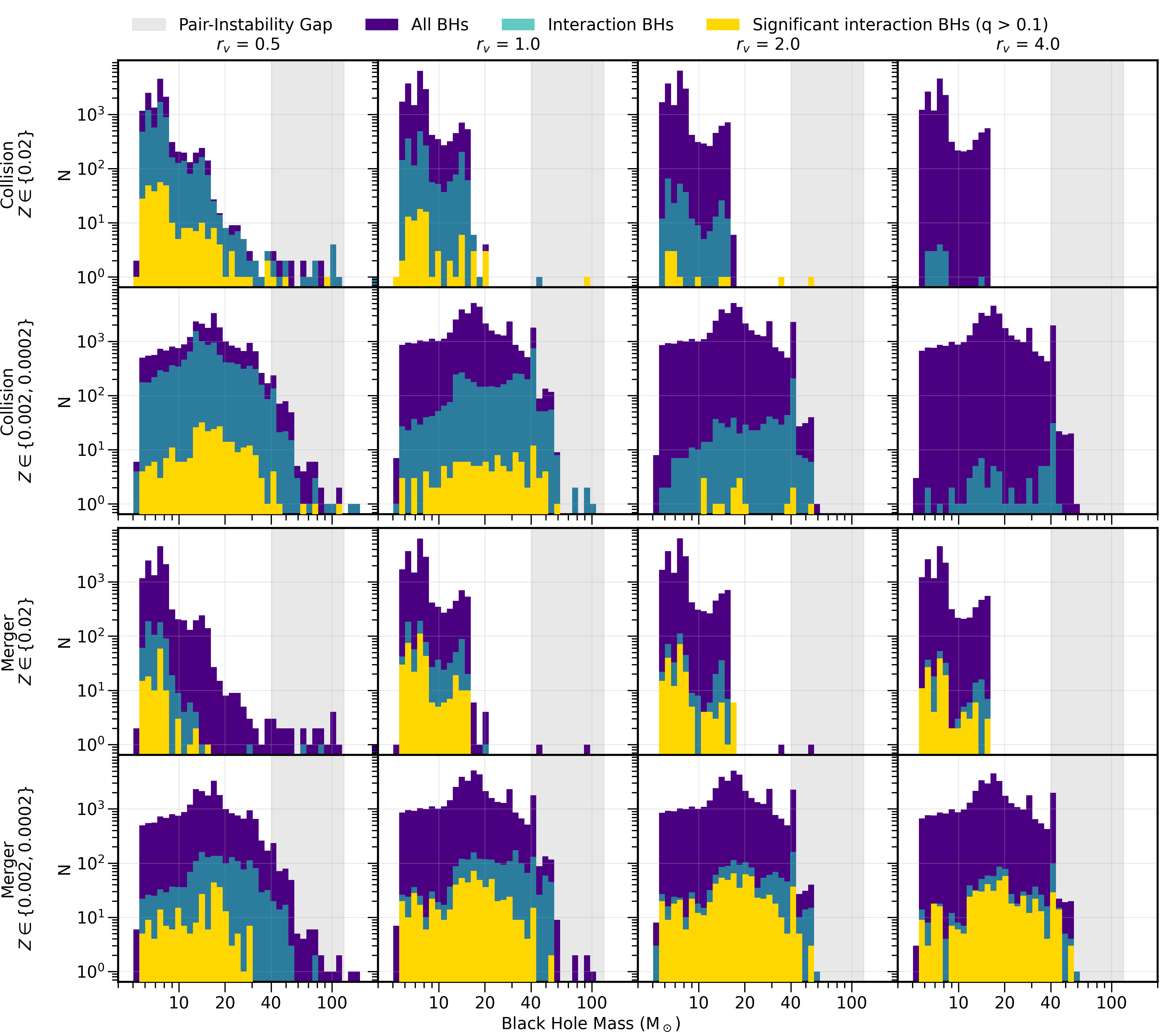}
    \caption{Mass distributions for all BHs formed via stellar collapse in our \texttt{CMC Cluster Catalog} models. The $4 \times 4$ grid is organized by formation channel (top two rows: Collisions; bottom two rows: Mergers) and progenitor metallicity (Rows 1 \& 3: $Z = 0.02$; Rows 2 \& 4: $Z \in \{0.002, 0.0002\}$). Columns represent the model virial radius ($r_v$), serving as a proxy for cluster density. Within each panel, distributions are shown for the total BH population (indigo), all stellar interaction products (teal), and "significant" interaction products with a mass ratio $q \ge 0.1$ (gold). Note that while the total number of BHs remains relatively consistent across $r_v$, significant interaction products are most prominent in the densest models ($r_v \leq 1.0\,$pc). Furthermore, lower metallicity models consistently produce a population of more massive BHs due to reduced mass loss from stellar winds. Our assumed boundary for the pair-instability mass gap is shown as a gray band in each panel.}
    \label{fig:BH_collisions_catalog}
\end{figure*}

We find MS+giant encounters comprise 85\% of the sample that ultimately form BHs. Motivated by this result, we next assess the impact of MS+giant interactions on the BH mass function. Figure~\ref{fig:BH_collisions_catalog} shows the mass function for all BHs formed via stellar collapse (this does not include second-generation BHs formed via BBH mergers), categorized by the initial cluster virial radius $r_v$ (four columns, from left to right) and metallicity $Z$ (sorted into rows of low $Z=[0.01,0.1]\times Z_{\odot}$ and high $Z=Z_\odot$). The top group of eight panels shows all BHs formed via MS+giant collisions, and the bottom group of eight shows those formed via MS+giant mergers. Each subpanel contains all \texttt{CMC Cluster Catalog} models corresponding to a given $[r_v,Z]$ combination; this includes 4 different cluster $N=[2,4,8,16]\times 10^5$ and 3 different galactocentric positions $R_{\rm gc}=[2,8,20]\,$kpc, or 12 models per subpanel in the high-metallicity group and 24 models per subpanel for the low-metallicity groups which includes both $Z=[0.01,0.1]\times Z_{\odot}$.

The various subpanels are then divided into three subsets: all BHs (indigo), all BHs formed via stellar collision/merger regardless of mass ratio (cyan), and BHs formed through significant ($q \geq 0.1$) collisions/mergers (yellow). For all $r_v$ combined, we see that each cut from indigo to cyan to yellow corresponds to roughly an order of magnitude decrease in sample size. This indicates that, in total, roughly 10\% of all BHs originate from collisions/mergers and roughly 10\% of these are significant. %However, when focusing on only those BHs above $40.5\,M_\odot$ (our assumed boundary for the pair-instability mass gap), we find that approximately a third ($27.6\%$) of BHs are formed via stellar collision/merger. This demonstrates that stellar collisions are a primary and efficient pathway for populating the upper BH mass gap \citep[e.g.,][]{Kremer2020}.

For the stellar collisions (top group of panels), we can see that as $r_v$ decreases, a much larger fraction of BHs are formed via collisions (significant or otherwise). For $r_v=0.5\,$pc, $44.2\%$ of BHs are formed via any type of stellar collision, and $1.5\%$ are formed from significant collisions. At $r_v=4\,$pc, these percentages decrease to $0.2\%$ and $0\%$, respectively. This trend arises simply because smaller $r_v$ correspond to higher cluster densities, and therefore higher collision rates (see Equation~\ref{eq:gamma}). This general trend holds independent of cluster metallicity. However, for stellar \textit{mergers} (bottom panels), this trend with $r_v$ is much less prominent. This is because stellar mergers largely arise from the primordial binaries in the cluster, which are constant across all models shown here. In this case, increasing the cluster density does not make a major difference. Last, Figure~\ref{fig:BH_collisions_catalog} shows that the low-metallicity groups produce significantly more massive BHs across all virial radii and interaction types, a direct consequence of our assumed stellar-wind prescriptions \citep[e.g.,][]{Agrawal2026}. However, the fraction of BHs formed via mergers/collisions does not appreciably change with metallicity.

\subsection{Implications for black hole spins}
\label{sec:CMCcatalog_spins}
 
\begin{figure*}
    \centering
    \includegraphics[width=0.95\linewidth]{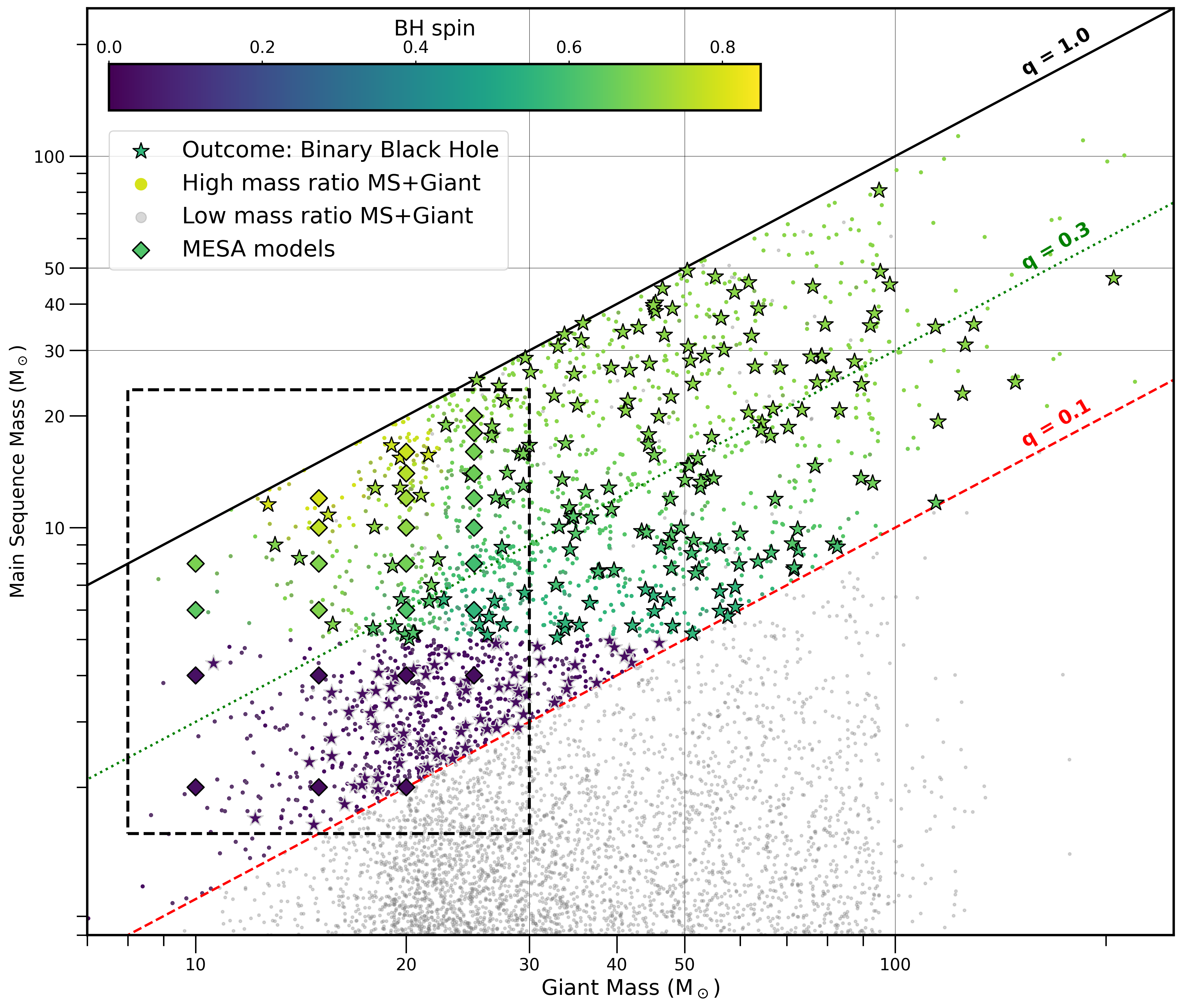}
    \caption{Stellar masses for all MS+giant interactions that ultimately collapse to BHs in the \texttt{CMC Cluster Catalog} models. We compare to the parameter space covered by our \texttt{MESA} models (shown as diamonds; see \citet{Tsuna2025} and Appendix). All scatter points are colored by the BH spin estimated for the interaction product, with spin values inferred from the \revision{nearest-neighbor \texttt{MESA} model.} The three diagonal lines denote mass ratios $q=[0.1,0.3,1]$. Based on the \texttt{MESA} models, we assume stellar interactions with mass ratio $q<0.1$ do not lead to significant spin up; we mark all these lower mass ratio events as gray scatter points in the bottom right of the plot. Stars denote BHs that go on to form BBH mergers later in the dynamical evolution of their host cluster. These are candidates for forming LVK event with non-zero component spins.}
    \label{fig:Catalog_MESA_comparison}
\end{figure*} 

The previous subsections demonstrate that stellar interactions are a common feature in the evolution of BH progenitors. In this subsection, we explore how these stellar collisions and mergers may imprint themselves upon the spin of the BH ultimately formed.

To estimate BH spins, we apply the recent model of \citet{Tsuna2025}, which performed a suite of stellar evolution calculations of post-main sequence binary mergers, with a metallicity of $0.5\,Z_\odot$, up to core-collapse using the stellar evolution code \texttt{MESA} \citep{Paxton11,Paxton13,Paxton15,Paxton18,Paxton19,Jermyn23}. The key finding of \cite{Tsuna2025} is that when more mass is gained in a merger, post-MS expansion is suppressed, and the star dies as a compact blue supergiant (BSG) rather than as a red supergiant (RSG) \citep[see also][]{Schneider24}. Upon failed supernovae, the rotating envelope of a compact BSG can mostly fall back instead of being ejected \citep{Fernandez18,Tsuna20,Ivanov21}, imparting a moderate to rapid spin for the final BH.
Qualitatively, the merger simulations of \citet{Tsuna2025} resemble the MS+giant interactions identified here as the most common type of pre-collapse stellar interaction in dense stellar clusters.

In this work, we run additional \texttt{MESA} models with a lower metallicity of $0.1\,Z_\odot$, closer to the characteristic value for globular clusters \citep[e.g.,][]{Harris1996}. We refer to Appendix A for details of the stellar models. We note that the upper limit of $M_1$ (giant mass) in our \texttt{MESA} grid is $25\,M_\odot$. Larger masses encounter numerical difficulties as the luminosity of the merger product approaches close to the Eddington limit, a well-known problem when simulating very massive stars \citep[e.g.,][]{Paxton13}. One may thus reasonably interpret our comparison of the \citet{Tsuna2025} results and our \texttt{CMC models} as a \textit{lower limit} on the true number of spinning BHs formed via stellar interactions, as we cannot make confident statements concerning the spin of the most massive stellar interactions. Note that roughly $50\%$ of our post-interaction BHs featured a merger involving a giant with mass larger than $25\,M_\odot$, so this this cut is unlikely to affect our main conclusions at more than a factor of order unity level.

To calculate the BH masses and spins from BSG progenitors (which have effective temperatures of $T_{\rm eff}>10^{3.9}\,$K at carbon depletion), we adopt the method of \cite{Tsuna2025} that solved the time-dependent fallback accretion (and outflow due to its super-Eddington nature) of the outer rotating envelope onto the newborn BH. For more extended stars with convective envelopes, we expect the angular momentum barrier of the infalling envelope due to its random convective motion leads to an explosion with an energy budget of $10^{48}$--$10^{49}$ erg \citep{Antoni23}, likely unbinding most of the loosely bound hydrogen-rich envelope. For the stars with $T_{\rm eff}<10^{3.9}$ K, we calculate the mass and spin of the BH assuming only the helium core (defined by hydrogen mass fraction $<0.1$) collapses, conserving its (rotational) angular momentum. The calculated BH spins for these non-BSG stars are small ($\lesssim 0.03$, as also seen in \citealt{FullerMa2019}), and nearly insensitive to the choice of the hydrogen mass fraction threshold.

\begin{figure*}
    \centering
    \includegraphics[width=1\linewidth]{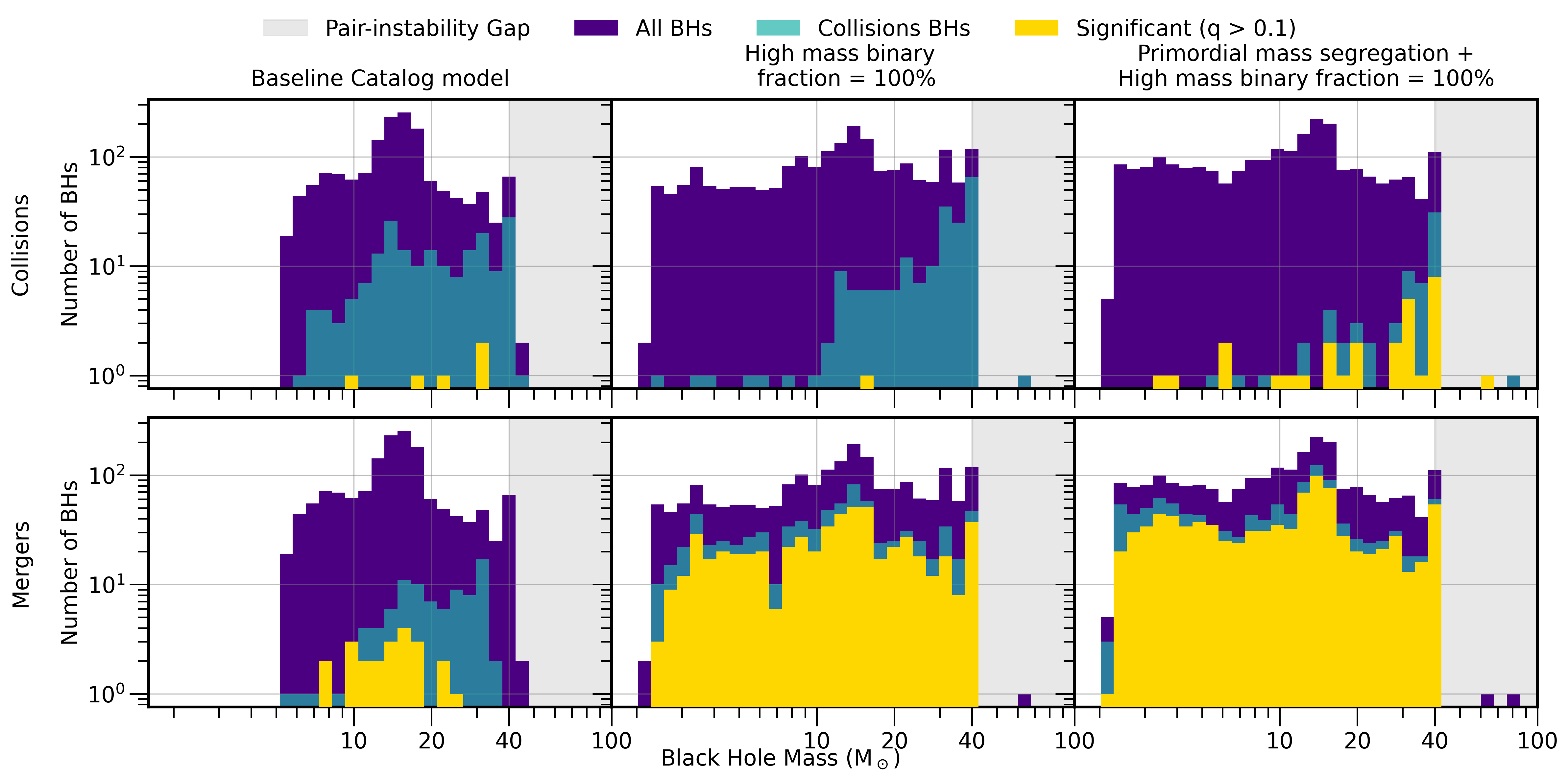}
\caption{BH mass distributions across three of our six cluster model variations. Columns represent varying initial conditions: Baseline \texttt{CMC Catalog} model (left), the high mass binary fraction of 100\% model (middle), and primordial mass segregation with 100\% binary fraction (right). The upper (lower) panels display BHs formed from stellar collisions (mergers). Indigo distributions represent the total BH population, teal indicates any interaction product, and gold highlights ``significant" interactions ($q \ge 0.1$). %Note that while high binary fractions and mass segregation significantly increase the total BH count and baseline interaction rates, the distribution of significant ($q \ge 0.1$) collision products remains qualitatively similar across the enhanced models.
}
    \label{fig:BHmasses_newmodels}
\end{figure*}

In Figure~\ref{fig:Catalog_MESA_comparison} we show as diamonds the MS mass versus giant mass for all $0.1\,Z_\odot$ \texttt{MESA} models computed here. Points are colored by the final spin of the BH estimated (see Figure~\ref{fig:BSGs_mass_vs_spin} in the Appendix). As shown, the mass ratio between the companion and the giant ($M_2/M_{\mathrm{giant}}$) is the primary determinant of the post-interaction outcome: when $M_2/M_{\mathrm{giant}} \gtrsim 0.3$ (indicated by the green dotted line), the collision deposits enough mass and angular momentum to suppress the giant’s post-main-sequence expansion, producing a BSG with a rapidly rotating envelope \citep[for more detail, see][]{Tsuna2025}.

We show as small scatter points (circles and stars) all MS+giant interactions in our \texttt{CMC Catalog} models that lead to BH formation. Colored points denote interactions with $q\geq 0.1$ (gold histograms of Figure~\ref{fig:BH_collisions_catalog}), and gray points denote lower mass ratios (teal histograms of Figure~\ref{fig:BH_collisions_catalog}). \revision{Points for $q\geq 0.1$ are colored according to the spin estimated by taking the nearest neighbor in the giant mass and main-sequence mass space from our \texttt{MESA} models (diamonds).} Circles show all BHs, while stars show those BHs that later pair up dynamically with another BH and undergo a BBH merger. We identify a total of $24,706$ MS+giant interactions of interest within our \texttt{CMC Catalog} models. Of these, $2,543$ are classified as ``significant'' with mass ratio $q \geq 0.1$. Of these, $1,132$ (roughly $44\%$) undergo a subsequent minor collision ($q < 0.1$) following the primary interaction. For simplicity, in cases where a multiple collision sequence occurs, we consider only the most significant merger/collision (defined as the one with mass ratio closest to unity) in the sequence. 

Among these significant interactions, $929$ have a mass ratio $q > 0.3$. Based on our \texttt{MESA} models, this value represents the critical threshold beyond which the collision product transitions from a slowly rotating red supergiant to a rapidly rotating blue supergiant. To ensure robust comparison, we must restrict our analysis to the parameter space directly bounded by the \texttt{MESA} grid (giant masses of $8-30\,M_\odot$ and MS masses of $1.5-23.5\,M_\odot$; black box in the figure). Within this restricted region, we find $496$ interactions that satisfy the $q > 0.3$ criterion.

We assume all of these $496$ are strong candidates for formation of a spinning BH via the processes described in \citet{Tsuna2025}. Of these spinning BH candidates, $30$ later undergo a BBH merger following dynamically pairing with other BHs in the cluster (compared to $8,058$ total BBH mergers occurring in the full \texttt{CMC Cluster Catalog}). Admittedly, this fraction alone is not particularly noteworthy, however in Section~\ref{sec:theoretical_uncertainties} we demonstrate how this fraction may increase significantly for variations in cluster properties.

\vspace{-0.5cm}

\section{Exploring model uncertainties}
\label{sec:theoretical_uncertainties}

\begin{figure*}
    \centering
    \includegraphics[width=0.95 \linewidth]{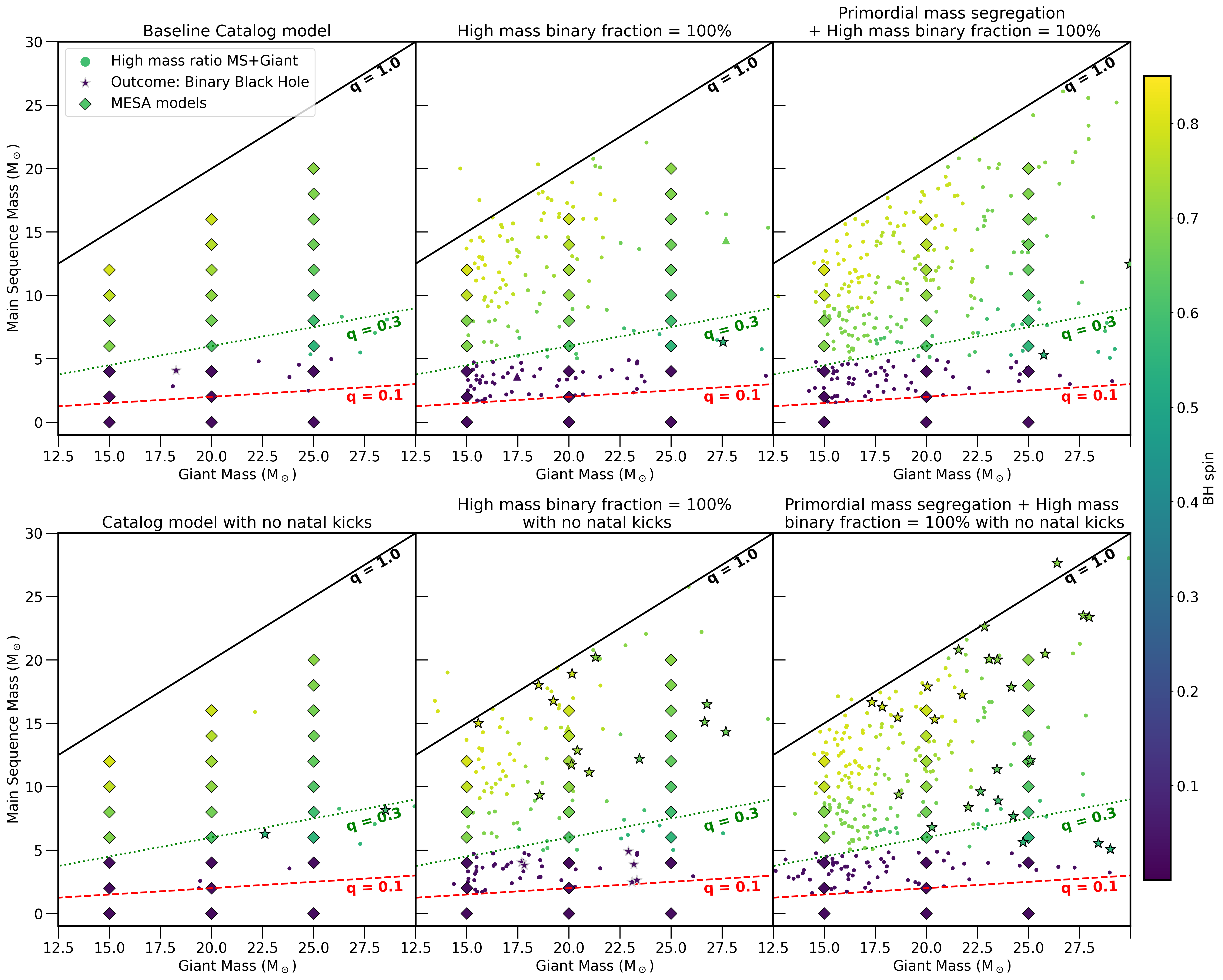}
    \caption{Stellar masses for all MS+giant interactions that collapse to BHs in our new models with increased binary fraction and primordial mass segregation (see Table~\ref{table:models}). As a baseline, we also show in the first column our results for the single analogous \texttt{CMC Cluster Catalog} model. We again compare to our \texttt{MESA} models, shown as diamonds. As in Figure~\ref{fig:Catalog_MESA_comparison}, all scatter points are colored by the estimated BH spin for the collision products. Stars denote BHs that go on to form BBH mergers later in the dynamical evolution of their host cluster. As shown, these new models increase the stellar interaction history for BH progenitors by up to a factor of ten, therefore increasing the likelihood for producing BHs and BBH mergers with non-zero spins.}
    \label{fig:MESA_comparison_newmodels}
\end{figure*}

%%%%%% Table goes here... %%%%%%%%%%%%%%%

\begin{table*}
	\centering
    \renewcommand{\arraystretch}{1}
    \tabcolsep=7pt
	\caption{Summary of BH and BBH merger formation statistics across simulated cluster models. The data categorizes total BH and BBH populations by their progenitor formation channels, distinguishing between collisions and mergers. While the initial total columns reflect all formation pathways, all subsequent interaction columns specifically isolate MS+Giant encounters, which dominate the BH formation channels. ``Significant" MS+Giant interactions denote events with a mass ratio $q \ge 0.1$. The $q \ge 0.3$ threshold isolated for interactions meeting the blue supergiant criteria of our \texttt{MESA} models, expected to be necessary to produce high-spin BHs. The ``MESA" columns restrict these $q \ge 0.3$ candidates to interaction products that fall strictly within our MESA grid mass constraints (MS: $1.5-23.5\,M_{\odot}$, Giant: $8-30\,M_{\odot}$; black box in Figure~\ref{fig:Catalog_MESA_comparison}). Finally, the 2G column reports the number of second generation BBH systems where one or more of the component BHs has already undergone a previous BBH merger.}
	\label{table:models}
	\begin{tabular}{lccc|cccc|ccccc}
\hline
\hline
 & \multicolumn{3}{c}{} & \multicolumn{4}{c}{Number of BHs} & \multicolumn{5}{c}{Number of BBH mergers} \\
 \cmidrule(r){2-4}
\cmidrule(lr){5-8}
\cmidrule(l){9-13}
Model & \multicolumn{1}{c}{$f_b$} & mass seg. & Kicks & \multicolumn{1}{c}{total} & \multicolumn{3}{c}{collision (merger)} & total & \multicolumn{3}{c}{collision (merger)} & 2G \\
\cmidrule(lr){6-8}
\cmidrule(l){10-12}
 & \multicolumn{3}{c}{} & \multicolumn{1}{c}{}{} & $q>0.1$ & $q>0.3$ & \multicolumn{1}{c}{MESA} &  & $q>0.1$ & $q>0.3$ & MESA & \\
\hline
    1 & 100\% & No & Stand. & 2279 & 5 (174) & 3 (102) & 0 (97) & 227 & 1 (5) & 1 (2) & 0 (0) & 14 \\
    2 & 100\% & Yes & Stand. & 2551 & 15 (309) & 11 (222) & 1 (208) & 206 & 6 (4) & 6 (1) & 0 (1) & 7\\
    3 & 100\% & Yes & No & 2461 & 12 (319) & 7 (234) & 0 (217) & 262 & 5 (40) & 3 (34) & 0 (22) & 16\\
    4 & 100\% & No & No & 2138 & 12 (155) & 10 (92) & 0 (87) & 270 & 7 (25) & 6 (15) & 0 (13) & 18\\
    5 & 5\% & No & No & 1534 & 9 (8) & 4 (1) & 1 (1) & 227 & 1 (2) & 0 (0) & 0 (0) & 21\\
    6 & 5\% & No & Stand. & 1536 & 7 (12) & 4 (2) & 0 (1) & 52 & 1 (1) & 0 (0) & 0 (0) & 12\\
    \hline
    Catalog & 5\% & No & Stand. & 209048 & 1725 (818) & 711 (218) & 349(147) & 13142 & 233 (43) & 92 (12) & 28 (2) & 1209 \\
\hline
\hline
\end{tabular}
\end{table*}    

With a baseline in mind from our \texttt{CMC Cluster Catalog} results, we now explore how assumptions connected to three key physical processes may potentially enhance the massive star collision rate, and therefore the formation rate of highly-spinning BHs. To reiterate the discussion from Section~\ref{sec:methods}, we increase the initial binary fraction for high mass stars ($>15\,M_\odot$) to $100\%$, we implement primordial mass segregation in our models, and we also assume all BHs are born without natal kicks to maximize BH retention. We run five new models in total: a $2\times2=4$ model grid turning on/off primordial mass segregation and turning on/off BH natal kicks, as well as one re-run of the equivalent \texttt{CMC Catalog} models without BH natal kicks, for completeness. All new models assume an initial population of $N=8\times10^5$ objects and metallicity $Z=0.1Z_{\odot}$. We summarize our key results from our supplementary \texttt{CMC} simulations in Table~\ref{table:models}.

Figure~\ref{fig:BHmasses_newmodels} shows the mass distribution for all BHs formed in these new models, analogous to Figure~\ref{fig:BH_collisions_catalog}. The three columns show the different model assumptions: our default \texttt{Catalog} model (left), our new model with $100\%$ binary fraction for massive stars (middle), and our new model with both $100\%$ binary fraction and primordial mass segregation (right). From left to right, these models can be read as most conservative to most optimistic in terms of their ability to produce spinning BHs via pre-collapse stellar interaction. As in Figure~\ref{fig:BH_collisions_catalog}, we separate collisions and mergers into separate rows. \revision{For each of the three model variations, we also ran a version with no natal kicks to establish an extreme upper limit for black hole retention. While neglecting natal kicks does not impact the upper part of the retained black hole mass function, it does alter the lower-mass regime ($\leq 20 M_\odot$) due to the mass-dependent kick prescription of \citet{Fryer2012} Specifically, in the baseline \texttt{Catalog} model, 39\% of all black holes formed with $M\leq 20 M_\odot$ were ejected at birth, while 57\% of low-mass black holes from high-mass-ratio stellar interactions were ejected at birth. This difference does not impact our key conclusions at more than a factor order unity level.} As in Figure~\ref{fig:BH_collisions_catalog}, the blue histogram shows mass distribution for \textit{all} BHs formed in each model (most of which do not undergo stellar interactions), teal shows those BHs that form following at least one stellar interaction of any mass ratio, and gold shows the BHs that form via significant interactions with $q>0.1$. 

Figure~\ref{fig:BHmasses_newmodels} demonstrates a few key points. Relative to the baseline \texttt{CMC Catalog} models, we observe a substantial increase in total merger rates, which can be attributed primarily to the $100\%$ primordial binary fraction for massive stars. In the baseline catalog model (bottom left panel), only $6\%$ (93 out of 1528) of BHs are formed from stellar mergers. However, when the massive star binary fraction is increased (bottom middle panel), this fraction jumps to $39\%$ (796 out of 2046), and to $49\%$ (1167 out of 2359) when primordial mass segregation is also included (bottom right). Furthermore, the proportion of these mergers that are dynamically significant ($q \ge 0.1$) rises dramatically from $24\%$ in the baseline model to $70\%$ in the enhanced binary models. 

While the total number of BHs formed via significant mergers increases significantly as we increase the binary fraction, the number of BHs formed via significant collisions remains largely unchanged (compare gold histograms in top left and top middle panels). However, once we introduce primordial mass segregation (top right panel), the increased stellar density in the core enhances the collision rate, leading to more significant collisions (roughly $30\%$ of all collisions in this model have $q\geq 0.1$).

\revision{While primordial mass segregation enhances the core density, it counter-intuitively decreases the total number of collisions that result in black hole formation as seen in the top middle and top right panels of Figure~\ref{fig:BHmasses_newmodels}. In the unsegregated model, massive stars are distributed more uniformly and undergo frequent collisions with more abundant lower-mass stars. Conversely, primordial mass segregation concentrates massive stars in the core where they preferentially collide with each other. This reduces the overall collision count but shifts the distribution toward high-mass-ratio events, as the massive stars are effectively separated from the broader low-mass population.}

In Figure~\ref{fig:MESA_comparison_newmodels}, we present an analog to Figure~\ref{fig:Catalog_MESA_comparison}, applied across our new models. Here we zoom in on the overlap region between our \texttt{CMC} interactions and our \texttt{MESA} models (boxed region of Figure~\ref{fig:Catalog_MESA_comparison}). As in Figure~\ref{fig:BHmasses_newmodels}, the different panels highlight the various initial cluster conditions. The top (bottom) row displays models with standard natal BH kicks (no natal kicks).

In the standard \texttt{CMC Catalog} models (first column), we observe only a single notable post-interaction candidate, with the vast majority of post-interaction BHs falling outside the \texttt{MESA} bounds or below the $q>0.3$ mass-ratio threshold. However, implementing a higher binary fraction (second column) results in a sharp increase, yielding 97 post-interaction candidates. The addition of primordial mass segregation (third column) further boosts the count to 208 candidates. Finally, the no-kick models (bottom row) demonstrate a substantial increase in the number of post-interaction BHs that later undergo BBH mergers. Specifically, the number of BBH mergers (denoted by stars) jumps from 0 to 13 for the high-binary-fraction model, and from 1 to 22 when combining primordial mass segregation with the high binary fraction.

\vspace{-0.5cm}

\section{Implications for gravitational wave recoil kicks}
\label{sec:GWrecoil}

\begin{figure*}
    \centering
    \includegraphics[width=1\linewidth]{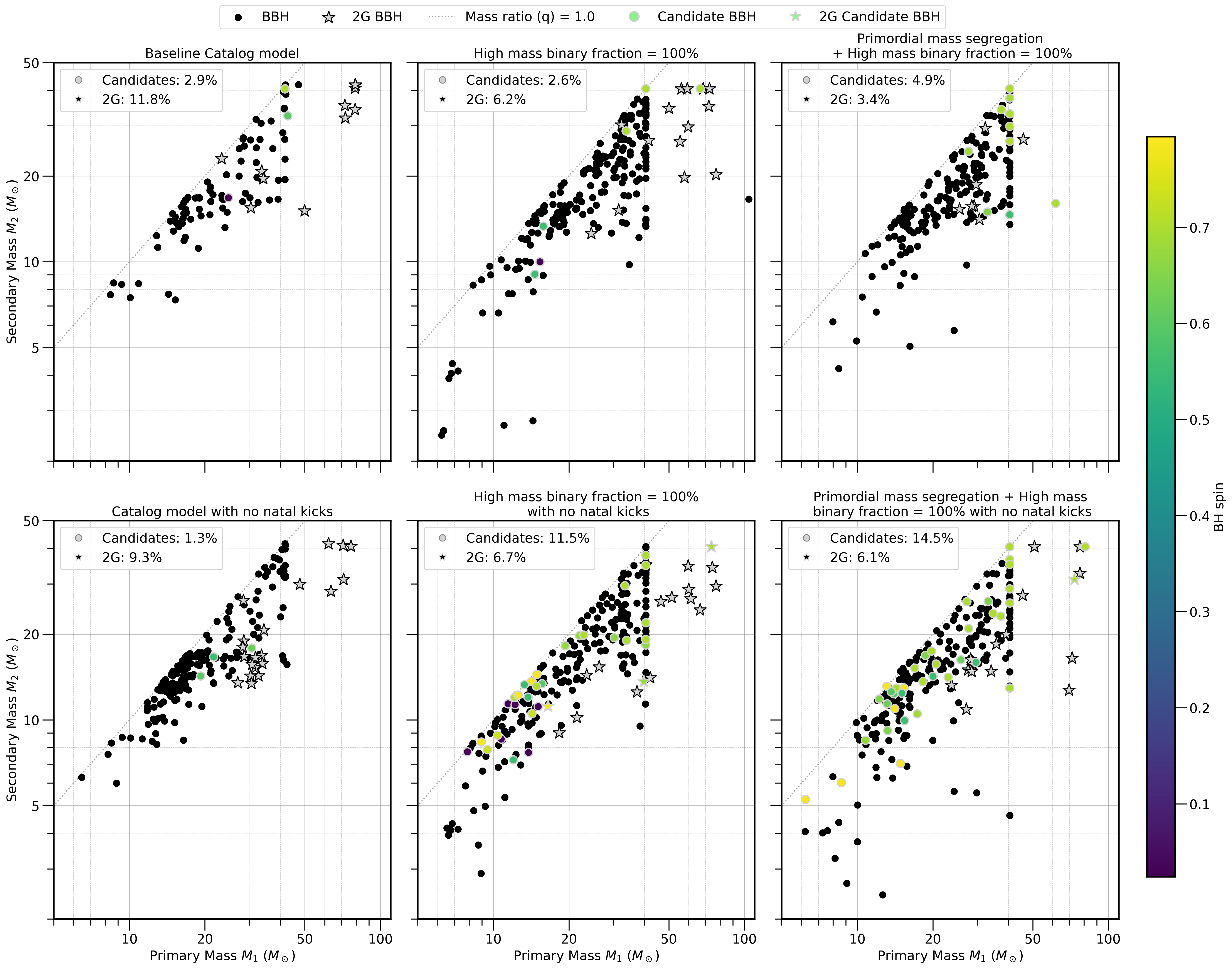}
    \caption{Secondary versus primary mass for all BBH mergers occurring in the models presented in Table~\ref{table:models}. Black points mark first generation (1G) mergers that did not undergo stellar collisions/mergers, i.e., ``pristine'' dynamical mergers. \revision{Light gray} stars mark BH mergers with at least one second generation component born via a previous BH merger. Colored circles (labeled ``Candidates'') mark the mergers with at least one component born with non-zero natal spin as a result of a pre-collapse stellar merger/collision. The color denotes our estimated BH spin, based on the \texttt{MESA} results. In our most optimistic models, we find that the number of spinning BH mergers from stellar interactions is comparable to those from second-generation mergers.}
    \label{fig:BBHmergers}
\end{figure*}

\begin{figure*}
    \centering
    \includegraphics[width=0.6\linewidth]{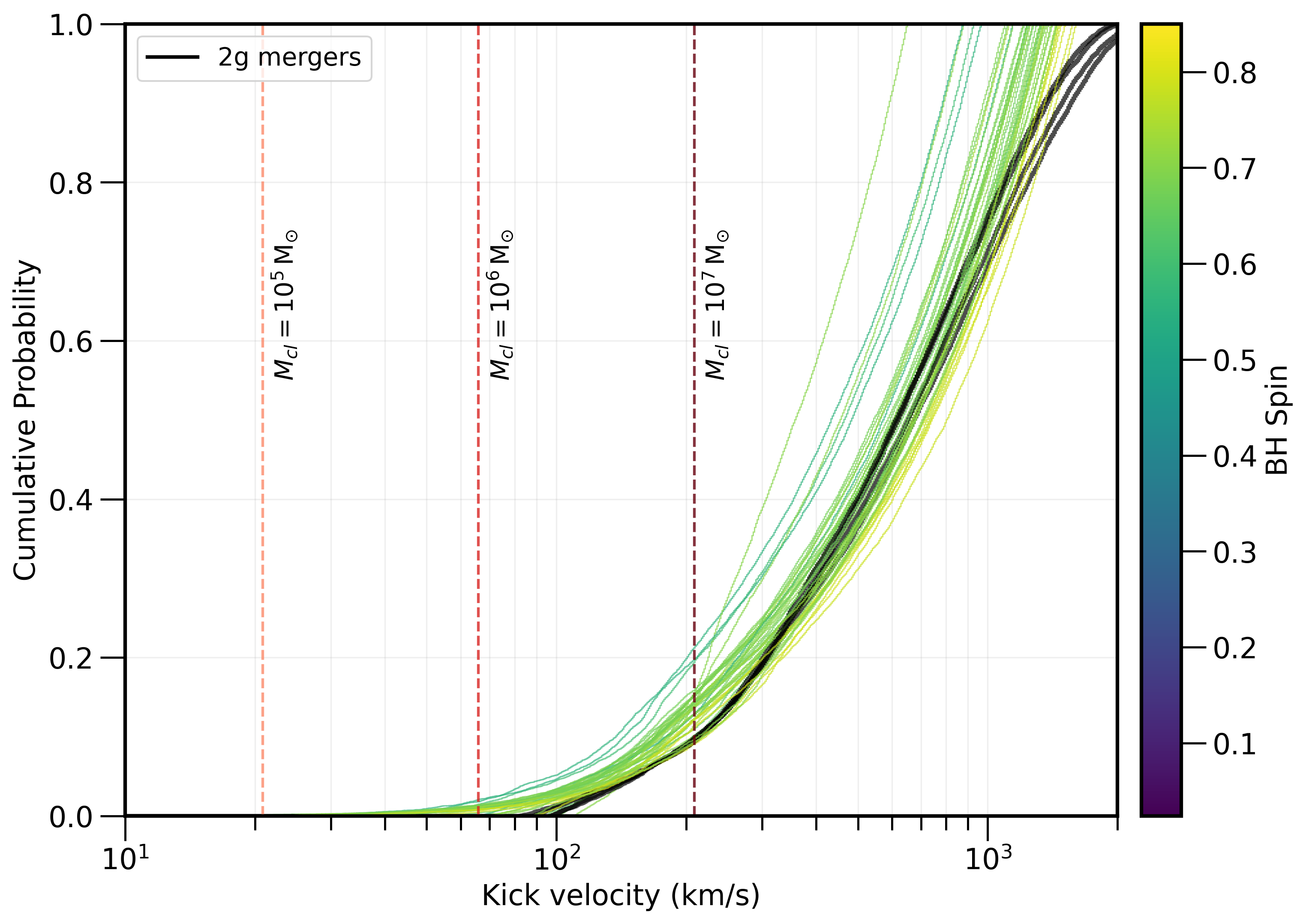}
    \caption{Cumulative probability distribution of gravitational-wave recoil velocities for all stellar interaction products that eventually participate in BBH mergers in the \texttt{CMC} models of Table~\ref{table:models}. %Recoil velocities are calculated utilizing the component masses from \texttt{CMC}, the specific spins predicted by our \texttt{MESA} models, and by performing $10^3$ random angle draws for spin and orbit vectors to create the distribution.
    Colored curves shows recoil kicks for BBH mergers involving a post-interaction BH and a ``normal'' 1G BH (color denotes the predicted spin for the post-interaction component), and black curves denote mergers involving a second generation component. Vertical lines represent characteristic cluster escape velocities for three different cluster masses, serving as a reference for the likelihood of merger product retention. The key takeaway is that a subpopulation of spinning BHs born following stellar interactions inhibits post-merger retention and therefore suppresses hierarchical mergers.}
    \label{fig:all_gw_kicks}
\end{figure*} 

When a spinning BH formed via a pre-collapse stellar interaction subsequently participates in a BBH merger, its high spin induces a much stronger gravitational-wave recoil kick. This in turn, makes retention of the merger remnant within the host cluster more challenging. We explore these implications in this section.

In Figure~\ref{fig:BBHmergers}, we show secondary mass versus primary mass for all BBH mergers formed across our new suite of six models. Mergers of ``vanilla'' first generation BBHs are denoted by black circles. Second-generation (2G) mergers where at least one component is the remnant of a previous BBH merger are marked as light \revision{gray} stars. Mergers involving at least one post-stellar-interaction BH (products of MS+G interactions with $q\geq 0.3$) are plotted as stars colored according to their predicted spin derived from our \texttt{MESA} models.

In the baseline \texttt{Catalog} model (top \revision{left}), mergers involving a post-interaction BH represent roughly $3\%$ of the total BBH population. This fraction increases marginally (up to roughly $5\%$) when we increase the binary fraction and add mass segregation (top-right panel). The most dramatic shift is observed in models without BH natal kicks (bottom panels). When all BHs are retained after formation \citep[even lower-mass BHs that are most likely to be kicked out of their host using the kick prescription of][]{Fryer2012}, this increases substantially the number of post-interaction BHs that can potentially merge later on. In the most optimistic environment (bottom-right), BHs spun up via stellar interactions compose roughly $15\%$ of the BBH merger population, exceeding the number of 2G BBH mergers.

To evaluate the implications for retention of these post-interaction BBHs with non-zero spins, we examine the gravitational-wave recoil velocities imparted to the final remnants. Figure \ref{fig:all_gw_kicks} consolidates the cumulative probability distributions of these recoil velocities for all post-interaction products that undergo BBH mergers later on. To illustrate the broader population trends, we include here every potential candidate from our full suite of new models in Table~\ref{table:models} (all scatter points with $q>0.3$ in Figure~\ref{fig:MESA_comparison_newmodels}).
Each curve is computed in post-processing from our \texttt{CMC} models, utilizing the specific spin values predicted by our \texttt{MESA} interpolation, the specific masses (obtained from \texttt{CMC}) of the BBH merger that occurs in the cluster models, and by performing $10^3$ random angle draws for the spin and orbit orientation. To compute recoil kick values, we use the fitting formulae outlined in \citet{GerosaKesden2016}. Within this population, we also identify post-interaction BHs that eventually merge with a second-generation (2G) BH, distinguished in Figure \ref{fig:all_gw_kicks} by thick black lines. Because 2G BHs possess their own significant spin magnitudes (typically $a\approx 0.7$ as discussed in Section~\ref{sec:introduction}), the resulting kick distributions for these systems often exhibit distinct, narrower profiles compared to the colored-curve cases where the post-interaction BH mergers with non-spinning 1G BH.

The escape velocity of a stellar cluster can be estimated roughly as
\begin{equation}
    \label{eq:v_esc}
    v_{\rm esc} \approx 66 \Bigg( \frac{M_{\rm cl}}{10^6\,M_\odot} \Bigg)^{1/2} \Bigg( \frac{r_h}{4\,\mathrm{pc}} \Bigg)^{-1/2}\,\mathrm{km/s},
\end{equation}
where $M_{\rm cl}$ is the total cluster mass and $r_h$ is the half-mass radius \citep[e.g.,][]{BinneyTremaine2008}. For reference, we also show in Figure~\ref{fig:all_gw_kicks} three vertical dashed lines corresponding to escape velocities for cluster mass $M_{\rm cl}/M_\odot = [10^5,10^6,10^7]$ and fixed $r_h=4\,$pc (assuming that $r_h$ is weakly correlated with cluster mass).

This complete landscape reveals several trends. First, stellar-interaction-induced spin strongly limits remnant retention. In standard \texttt{CMC} simulations, many remnants of first generation mergers (1G+1G mergers) are retained under the default assumption of zero spin \citep[e.g.,][]{Rodriguez2019,kremer2020modeling}. However, when incorporating the spins predicted by our \texttt{MESA} models (typically $a \gtrsim 0.5$), the kick distributions broaden significantly and shift toward much higher velocities, meaning remnants that would have been retained are now likely to be ejected. For a fiducial cluster with $M_{\rm cl}=10^6\,M_{\odot}$ and $v_{\rm esc}=66\,$km/s, these results suggest that over $95\%$ of these post-interaction BBH mergers would escape via gravitational wave recoil.

\section{Discussion \& Conclusions} 
\label{sec:conclusions}

\subsection{Summary}

To summarize, we have identified the following key results:

\begin{enumerate}
    \item In dense stellar clusters, mergers and collisions are common features of the evolution of massive stars before they collapse into BHs. Using a large suite of cluster models performed using \texttt{CMC}, we find that up to nearly $50\%$ of all BHs may undergo at least one stellar merger/collision before formation, and up to roughly $10\%$ of BHs undergo a significant interaction with mass ratio $q>0.1$ that may potentially lead to efficient BH spin up. 

    \item The occurrence of these pre-collapse interactions depends significantly on assumptions about cluster initial conditions. We show the cluster virial radius $r_v$, primordial binary fraction, and the degree of primordial mass segregation all play a crucial role.

    \item Following \citet{Tsuna2025}, we simulate with \texttt{MESA} the mergers of massive giants and main sequence stars. By evolving these post-merger remnants to collapse, we estimate the spin of the BHs ultimately formed. We find mergers with mass ratio $M_{\rm MS}/M_{\rm Giant} > 0.3$ result in a blue supergiant, which ultimately result in BHs with dimensionless spin parameter $a \approx 0.5-0.8$. This critical mass ratio and the resulting BH spin is highly dependent on metallicity, with higher metallicity populations requiring higher mass ratios and resulting, in general, in lower BH spins.

    \item In our \texttt{CMC} model with $100\%$ binary fraction for high-mass stars, primordial mass segregation, and assuming BHs form with zero natal kicks, we find roughly $10\%$ of all BBH mergers feature at least one BH born with high spins through this process. This rivals the fraction of BBHs expected to contain a second generation BH born via previous BBH merger.

    \item The spin-up of post-interaction BHs broadens their gravitational-wave recoil kick distributions upon merger with other BHs. This makes the remnants of these mergers much more likely to be ejected from their host clusters relative to the standard assumption of zero natal spins. As a result, a subpopulation of non-zero spin BHs formed via stellar interactions may inhibit formation of hierarchical mergers in clusters.
    
\end{enumerate}

\subsection{Discussion \& future work}
\revision{While our results demonstrate that massive stellar mergers provide a viable pathway to highly spinning black holes, we emphasize that this is not the sole scenario; high natal spins can also be attained via hierarchical mergers, runaway stellar collisions, envelope collapse, accretion from a binary companion, and tidal spin-up of a progenitor by a companion star. Given our adoption of a 100\% binary fraction for massive stars in some models, these effects are expected to operate efficiently in our setups. Determining the relative contributions and observational signatures of our proposed mechanism compared to these diverse pathways remains an important goal for the field and }this study sets the stage for a number of areas of future work. We have focused here on analysis of stellar interactions in \texttt{CMC} models in post processing. Implementation of these effects into N-body codes like \texttt{CMC} are needed in order to self-consistently evaluate the long-term evolution of BH spins on BH merger retention.

Similarly, we have applied insight from a grid of \texttt{MESA} simulations of stellar interactions to estimate (also in post-processing) the post-merger properties of the BH progenitors. A more direct treatment of the merger (hydro)dynamics and subsequent stellar evolution would be ideal. For examples of this type of exercise, see recent work by \citet{Ballone2022,Costa2022}. For example, we have implemented BH natal kick prescriptions that are standard in the literature, yet are not self-consistently computed with the expected fallback and spin-up prescriptions suggested by our \texttt{MESA} models. We have attempted to bracket the range in uncertainties for BH natal kicks by exploring a range in possibilities. Ultimately, a self-consistent calculation of fallback, spin-up, and natal kick would be ideal, but of course numerical calculation of BH natal kicks is famously challenging and highly uncertain \citep[e.g.,][]{Chan2020,MandelMuller2020,Burrows2024,JankaKresse2024}.

We have focused primarily on the implication of BH spins on formation of gravitational wave sources, however these collapsar-like events that form spinning BHs are expected to also power bright electromagnetic transients \citep[for further detail, see][]{Tsuna2025}. This may include core-collapse supernovae similar to 1987A \citep[e.g.,][]{Menon2019}, long or ultra-long gamma-ray bursts \citep[e.g.,][]{Perna2018}, and fast luminous transients similar to the fast blue optical transients found in high-cadence optical surveys \citep[e.g.,][]{Drout2014}. Furthermore, the stellar collisions/mergers themselves (including the many lower-mass stellar interactions that do not lead to BH formation; see Figure~\ref{fig:all_collisions}) may produce interesting electromagnetic transients of their own, likely similar to luminous red novae \citep[e.g.,][]{MetzgerPejcha2017}. We reserve for future studies exploration of rates, the specific types of transients that may be powered, and how these may depend on stellar interaction details.

\section*{Acknowledgements}

\revision{We thank the anonymous referee for a constructive review of the paper.} D.T. is supported by Harvard University through the Institute for Theory and Computation Fellowship. F.K. acknowledges support from a CIERA Postdoctoral Fellowship. This research was supported in part through the computational resources and staff contributions provided by the Quest high-performance computing facility at Northwestern University.

%%%%%%%%%%%%%%%%%%%%%%%%%%%%%%%%%%%%%%%%%%%%%%%%%%
\vspace{-0.7cm}
\section*{Data Availability}

The data supporting this article are available on reasonable request to the corresponding author.


\vspace{-0.7cm}

\bibliographystyle{mnras}
\bibliography{mybib} % if your bibtex file is called example.bib

%%%%%%%%%%%%%%%%%%%%%%%%%%%%%%%%%%%%%%%%%%%%%%%%%%

%%%%%%%%%%%%%%%%% APPENDICES %%%%%%%%%%%%%%%%%%%%%

%%%%%%%%%%%%%%%%%%%%%%%%%%%%%%%%%%%%%%%%%%%%%%%%%%

\appendix

\begin{figure*}
    \centering
    \includegraphics[width=0.8\linewidth]{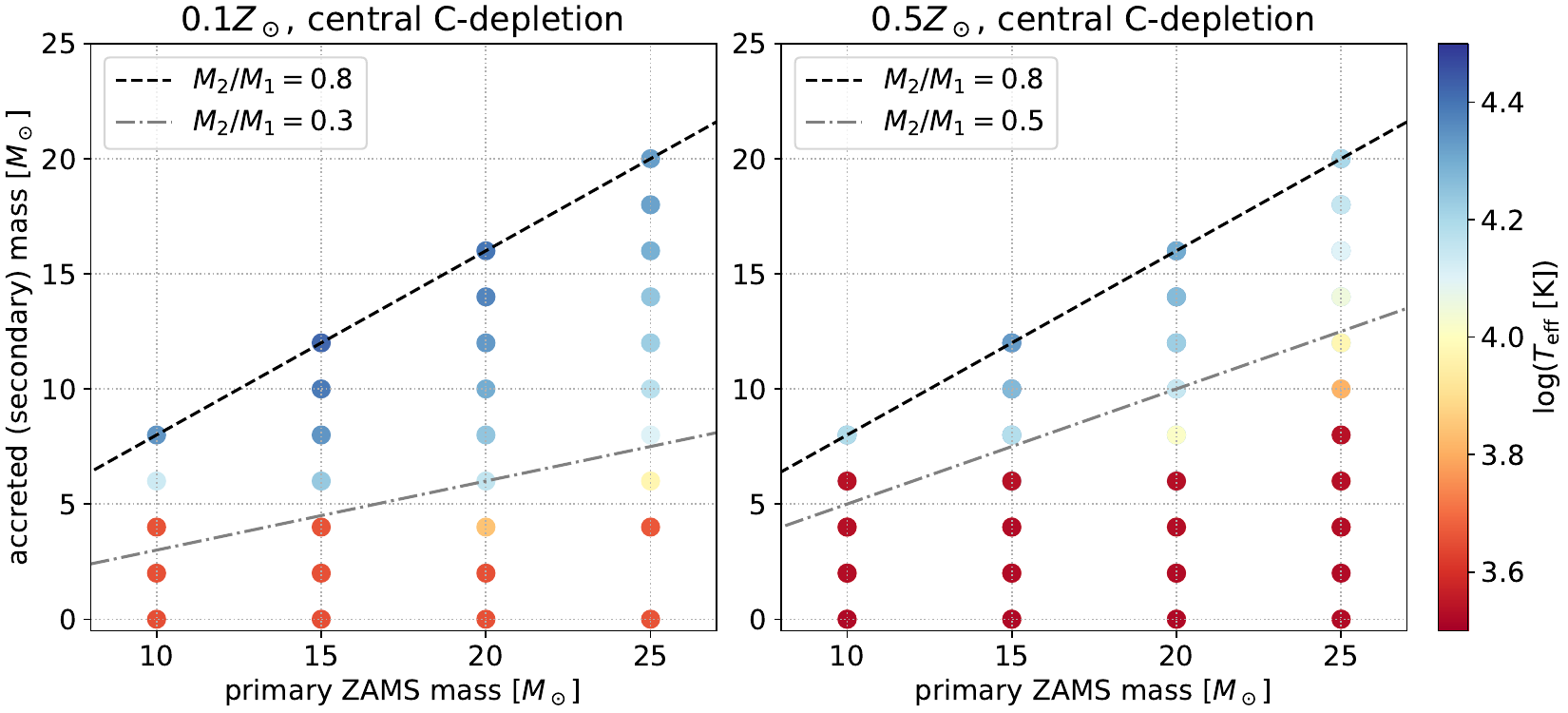}
    \caption{Effective temperatures of the \texttt{MESA} models at core carbon depletion near core-collapse. Left panels show our newly calculated models with $Z=0.1\,Z_\odot$, while right panels show the models with $Z=0.5\,Z_\odot$ of \citet{Tsuna2025}.}
    \label{fig:mesa_models}
\end{figure*}

\begin{figure*}
    \centering
\includegraphics[width=0.5\linewidth]{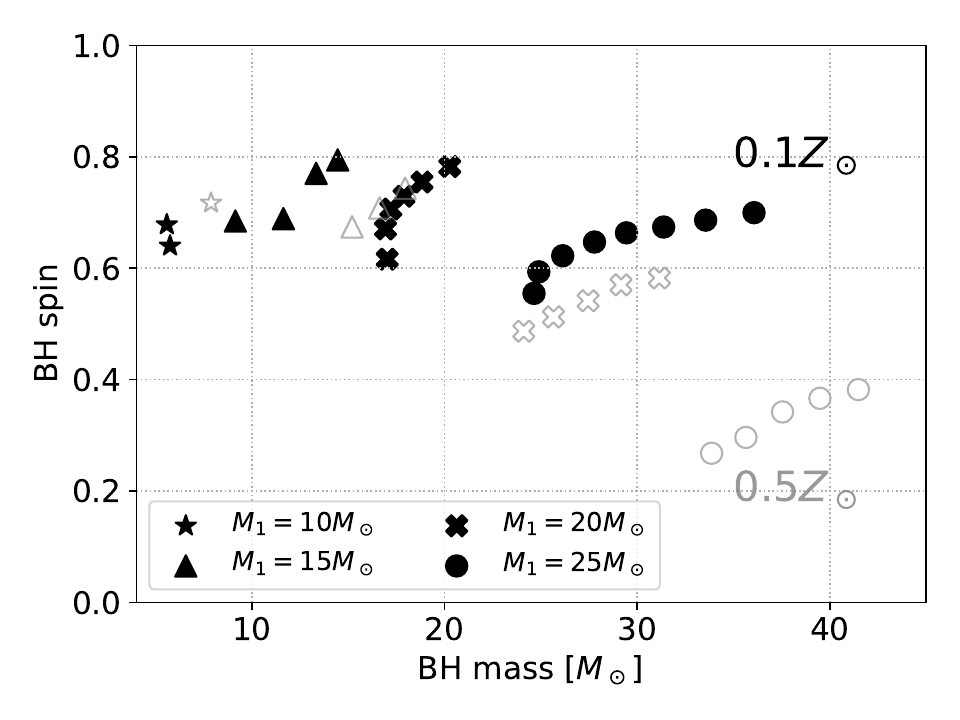}
    \caption{BH mass and spin predictions, for the \texttt{MESA} models that are BSGs at core carbon depletion. Black filled symbols are newly calculated models with $Z=0.1~Z_\odot$, while grey open symbols show models with $Z=0.5~Z_\odot$ of \citet{Tsuna2025}.}
    \label{fig:BSGs_mass_vs_spin}
\end{figure*}

\section{Description of the additional \texttt{MESA} models}
In this work we employ \texttt{MESA} as in \cite{Tsuna2025} to generate additional stellar models of the merger products with metallicity of $0.1Z_\odot$. We adopt a same grid of $(M_1, M_2)$, with $M_1=[10,15,20,25]~M_\odot$ and $M_2$ from $0$ (single star) to $0.8M_1$ with spacing of $2~M_\odot$. We also update the mixing length parameter governing convection from $\alpha_{\rm MLT}=1.5$ in \cite{Tsuna2025} to $\alpha_{\rm MLT}=3$, motivated from observational studies of local RSGs and simulations of RSG envelopes \citep{Chun18,Goldberg22} as well as for better convergence in the post-merger relaxation phase. This parameter mainly controls the surface properties for stars with convective envelopes (e.g., RSGs), but does not affect the stars with radiative envelopes (e.g., BSGs) of our main interest. The new \texttt{MESA} models are mostly calculated until core-collapse, except for one model of $(M_1,M_2)=(25~M_\odot, 2~M_\odot)$ that encountered numerical difficulties post merger.

Figure \ref{fig:mesa_models} shows the effective temperatures of the merger products at core carbon depletion, after which the stellar envelopes are not expected to evolve significantly until core collapse. Similar to the $Z=0.5Z_\odot$ grid in \cite{Tsuna2025}, there is a threshold on $M_2$ above which the merger products form BSGs at core-collapse. We find that across the range of $M_1$ considered, lower metallicity reduces the threshold of $M_2$ for the merger product to die as BSGs. A lower metallicity more likely leading to hotter BSGs is expected also in single star evolution, due to reduction in the radiative temperature gradient in the star via the lower opacity. While there is no clear mass ratio $M_2/M_1$ that separates from BSGs to RSGs, our model grid implies that the merger products could die as BSGs for  $M_2/M_1\gtrsim 0.3$ when $Z=0.1Z_\odot$, compared to $M_2/M_1\gtrsim 0.5$ for $Z=0.5Z_\odot$, as in \citet{Tsuna2025}.

Figure \ref{fig:BSGs_mass_vs_spin} shows the masses and spins of the BHs, for the merger product that end their lives as BSGs. The BH parameters are calculated by the prescription of \cite{Tsuna2025} (see their Section 4.2 and Figure 8), under the assumption that these progenitors undergo failed supernovae and the bulk of the envelope (and core) falls onto the newborn BH. Compared to the $0.5Z_\odot$ models, the $0.1Z_\odot$ models have higher BH spins and {\it lower} BH masses for the same $M_1$. This is because of the weaker stellar winds leading to more rapid progenitor rotation, which has two effects upon BH formation: (i) the angular momentum added to the newborn BH is larger, leading to a higher BH spin, and (ii) more infalling matter tends to circularize and form a disk instead of falling straight into the BH. Most of the disk material is eventually lost by a super-Eddington disk wind, therefore reducing the final mass of the BH.

% Don't change these lines
\bsp	% typesetting comment
\label{lastpage}
\end{document}